# Link creation and profile alignment in the aNobii social network

Luca Maria Aiello*, Alain Barrat†‡, Ciro Cattuto‡, Giancarlo Ruffo* and Rossano Schifanella*
*Università degli Studi di Torino, Computer Science Department, Torino, Italy
†Centre de Physique Théorique (CNRS UMR 6207), Marseille, France
‡Institute for Scientific Interchange (ISI) Foundation, Torino, Italy

*Abstract*—The present work investigates the structural and dynamical properties of aNobii[1], a social bookmarking system designed for readers and book lovers. Users of aNobii provide information about their library, reading interests and geographical location, and they can establish typed social links to other users. Here, we perform an in-depth analysis of the system's social network and its interplay with users' profiles. We describe the relation of geographic and interest-based factors to social linking. Furthermore, we perform a longitudinal analysis to investigate the interplay of profile similarity and link creation in the social network, with a focus on triangle closure. We report a reciprocal causal connection: profile similarity of users drives the subsequent closure in the social network and, reciprocally, closure in the social network induces subsequent profile alignment. Access to the dynamics of the social network also allows us to measure quantitative indicators of preferential linking.

## I. INTRODUCTION

Social media have become rich information ecosystems where the activity of users is entangled with the social networks that mediate the interaction of users with one another and with the content they process. Data from these systems expose a number of fundamental mechanisms that drive the dynamical evolution of on-line social networks. The present work focuses on the structural and dynamical properties of the social network of aNobii, an on-line social bookmarking system designed for reading lovers. aNobii allows users to share reviews of books they have read and to receive reading suggestions from other users. The main goal of this study is to shed light on the mechanism of online social linking, through investigations of the interplay between social links and profile similarity, of the topological features of new links, and of the effect that the creation of a link has on the properties of its endpoints.

To our knowledge, the aNobii social network has not been analyzed so far, and this paper represents the first contribution to a static and dynamic investigation on such a network. Moreover, aNobii presents some interesting points that deserve a thorough investigation. First of all, aNobii users largely provide details such as their geographical location (97% of users specify the country and 38% include also the city). This motivated us to analyze the influence of geographic parameters on social aggregation, and interest-based networking. Secondly, conversely to many other Online Social Networks (OSNs), users are linked to each other through two mutually-exclusive relations: *friendship* (established with people known in real life) and *neighborhood* (established with people with interesting libraries). Such links are directed, and they can be created without the explicit acceptance of the linked user. This makes aNobii a peculiar framework for performing a thorough study on the understanding of links formation, and how users behave and connect to each other. Finally, since our work is based on a set of snapshots of the system over time, we could investigate the time evolution of links, providing the opportunity of exploring the two-way causal relation between profile similarity – defined in terms of several features – and the creation of new social links.

In summary, our main contributions are the following. We introduce a new dataset made by fairly complete snapshots of the aNobii network, reporting a static analysis on its main properties and studying the relationship between network topology and users' features. We make dataset available upon request. We moreover mine the geographical information provided by users, exposing its role in social partner selection. Finally, we analyze the dynamics of link formation and profile alignment, and provide evidence for the entanglement of these phenomena in both friendship and interest networks.

The structure of the paper is as follows. Section II provides an overview of the relevant literature and state of the art. Sections III and IV describe the dataset used for the analysis and discuss the static properties of the users' social network and activity patterns. Section V reports the geographical features of the social network and relates social links with the underlying geographical distribution of users. Section VI investigates the role of profile similarity, 1) as a condition for the creation of a new social link, and 2) as the result of the creation of a social link. Section VII summarizes the contributions of this paper and points to a number of open questions.

## II. RELATED WORK

In recent years, many efforts have been spent towards the analysis of topological properties and of the evolution of OSNs. Analysis over time of large-scale social networks' topological features like diameter, clustering coefficient and mixing patterns are performed in [1], where structural differences between real life social network and OSN are explored, and in [2], where a fast and significant link reciprocation phenomenon is observed in directed social graphs. In [3] an interaction graph of Facebook determined by social links that are effectively exploited for user-to-user communication is extracted and compared with the full social network; structural properties of both graphs are inspected over time. The dynamics of a network of students, faculty and staff at a large university is studied in [4] focusing on the combination of effects arising from the network topology itself and the organizational structure in which the network is embedded.

---
[1] http://www.anobii.com/

Several other studies focus on link characterization, namely on the patterns that describe the creation of links and how social ties features evolve in time. In [5], the authors state that large OSNs are composed by three structural regions: the singleton nodes, the giant component, and a middle region formed by various isolated communities detached from the core. The authors investigate how small isles merge together or join with the giant component and propose a formal model to explain such network evolution. An evolutionary study of the Flickr OSN is performed in [6]. It shows that the link formation process is characterized by reciprocation and by a tendency to link users who are already close in the network. In [7] the evolution of the intensity of active social ties in terms of number wall posts is studied in Facebook activity network. Results show a general decreasing trend: on average, users interact less and less over time. In [8], the trust dynamics of CouchSurfing.com network are studied. Here, the friendship links are found to be the most predictive feature of whether a user will express his vouch for another in the reference section of the website. A high degree of reciprocation in vouching is found as well. While these works study the dynamics of links mainly from a topological point of view, in this paper we describe links creation and evolution in terms of the similarity between the features of the users which these links connect. This change of perspective allows us to infer interesting causal relationships between users' profiles and social activity.

Ref. [9] investigates the interplay between homophily-driven creation of social ties and the influence that neighbors exert on each other's behavior based on data from the active editors of the Wikipedia collaboration network. Editors tend to become aware of each other, and to establish direct communication, when they start having many editing activities in common and, on the other hand, the direct interaction between them is found to result in further alignment between their activities. Investigating the causal relation between homophily and neighbor selection is a goal of our work as well. However, the social network at hand is of a very different nature from Wikipedia, and our focus is on the analysis of the profile features, shared metadata, and topicality rather than on collaboration patterns.

Fewer works focus on how geographical aspects relate to dynamics of interaction in OSNs. In [10] an analysis on LiveJournal reveals an inverse linear relationship between the geographic distance and the probability of becoming friends. The authors report that most friendship links are derived from geographical processes. Similar results are shown in [11], where a Facebook-like OSN for German-speaking students is analyzed with respect to the geographic location of users. It is observed that the rate of acquaintanceship rapidly drops with the increase of the geographical distance between them. In our work we found that similar conclusions apply not only to geographically-driven networks (i.e. OSNs where links are preferentially established between individuals who know each other in real life, like Facebook) but also to pure interest networks, where social ties are created on an affinity basis.

III. DATASET DESCRIPTION

aNobii was created in Hong Kong in 2005 and soon became popular abroad, especially in Italy. Users in aNobii can insert information regarding a digital book collection and about themselves. Profile data include in particular country and home town. The book collection is articulated in two sections: a *wishlist*, which contains the book titles that the user has planned to read, and a *library* which is a list of book titles that the user has already read or she is currently reading. Users fill up their libraries and wishlists by selecting books from the aNobii database, which indexes about 20 millions different titles along with their metadata (such as author, publication year, etc.). Each book in the library can be annotated with keywords (*tags*), a rating, a review and a reading status (e.g. finished, reading, etc.). Users are connected to each other through two mutually-exclusive types of ties: *friendship* and *neighborhood* links. At aNobii's website suggestion, friendship should be established with people you know also in real life, while neighborhood is intended for people you do not know but whose library you consider interesting. Apart from this formal distinction, such links behave exactly in the same way: they are both *directed* and they enable the notification of the linked user's library updates. A tie can be established without the explicit consent of the linked user, which is not even informed when a new incoming link is created. Another social tool available is the affiliation to thematic *groups*, which are created by users and whose membership is publicly accessible.

We collected several snapshots of aNobii network through BFS crawling of neighborhood and friendship networks, starting from a random seed. Since the social relations are directed, we explored the giant strongly connected component and the full out component of the network. We extracted *users' profile data*, *library information* and *groups affiliations* through web scraping. We took six snapshots of the network, 15 days apart, starting from 11/09/2009. Anonymized versions of the snapshots are available upon request.

IV. STATIC ANALYSIS

In the following, we perform a topological analysis of the 12/24/2009 snapshot, which is used as a representative sample, for brevity. Topological features are stable over time: all the curves for different snapshots can be superimposed.

*A. Network characteristics*

Table I shows some basic properties of the friendship and neighborhood networks and of their union. Statistics show a general high level of link reciprocation, which suggests that users become aware of new incoming links, even if the system does not explicitly notify them. This is due to the *wall post* service offered by aNobii, which is often used by people for introductions to new friends or neighbors. Some structural differences between the two networks show that friendship and neighborhood ties are not used interchangeably by users. Neighborhood network is denser and with a higher degree centralization [12]; this reflects the fact that the interest in other users' readings tends to concentrate toward a core of 'hot' libraries which are monitored by many users. As a result of this feature, plus the fact that neighborhood spans a smaller fraction of users, neighborhood network's diameter (i.e. the maximum shortest path length) and average shortest path length are considerably smaller than the friendship network's ones. The union of the two networks has a diameter of 20 and an average shortest path length higher than 5, which are large values for a node set size smaller than 100k nodes, if compared to other well-known social networks (see [2]). These features suggests that the aNobii social network is characterized by an elongated shape, mainly caused by geographical aspects we discuss in Section V. Since there is no link overlap between friendship and neighborhood networks and since many of the results we obtained are independent from the network type, in this work we will mostly focus on the union network, which connects the whole set of users. In the following, we refer to the union network as the aNobii social network.

|  | Friendship | Neighborhood | Union |
|---|---|---|---|
| Nodes | 74,908 | 54,590 | 86,800 |
| Links | 268,655 | 429,482 | 697,910 |
| Reciprocation | 0.71 | 0.45 | 0.57 |
| $<k_{out}>$ | 3.6 | 7.9 | 8.0 |
| WCC size | 68,624 | 54,246 | 86,800 |
| SCC size | 46,253 | 29,110 | 62,195 |
| Density | $4.8 \cdot 10^{-5}$ | $1.4 \cdot 10^{-4}$ | $9.3 \cdot 10^{-5}$ |
| Average SPL | 7.3 | 4.7 | 5.3 |
| Diameter | 25 | 15 | 20 |
| Degree centr. | 0.0082 | 0.12 | 0.079 |

TABLE I
FRIENDSHIP, NEIGHBORHOOD AND FULL SOCIAL NETWORK STATISTICS (SPL=SHORTEST PATH LENGTH; WCC=WEAKLY CONNECTED COMPONENT; SCC=STRONGLY CONNECTED COMPONENT).

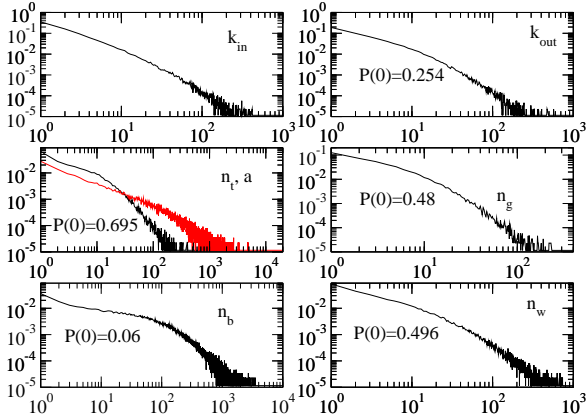

Fig. 1. Distributions of the measures of activity of aNobii users: in-degree $k_{in}$ and out-degree $k_{out}$ in the social network, number of distinct tags $n_t$ and total tagging activity $a$ (total number of tags in a user's page), number of group memberships $n_g$, number of books in a user library $n_b$ and in a user wishlist $n_w$.

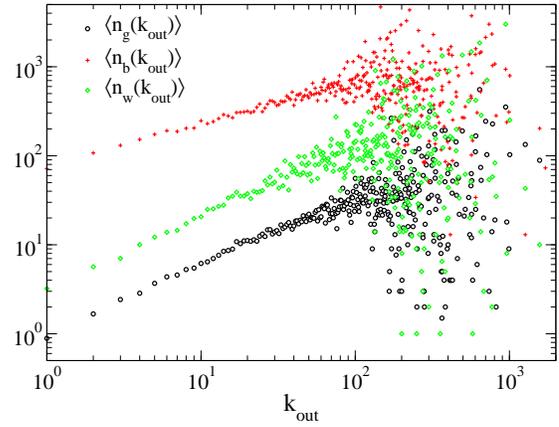

Fig. 2. Correlations between the activity of aNobii users and their number $k_{out}$ of out-links in the social network (number of declared friends and neighbors): average number of group memberships $\langle n_g \rangle$, of library $\langle n_b \rangle$ and $\langle n_w \rangle$ wishlist sizes, vs $k_{out}$.

|  | $k_{out}$ | $n_g$ | $n_b$ | $n_w$ |
|---|---|---|---|---|
| $k_{out}$ | 1 | 0.31 | 0.18 | 0.18 |
| $n_g$ |  | 1 | 0.32 | 0.31 |
| $n_b$ |  |  | 1 | 0.22 |

TABLE II
PEARSON COEFFICIENTS BETWEEN THE VARIOUS MEASURES OF ACTIVITY OF ANOBII USERS.

As described above, the activity of a user of aNobii has several aspects: exposing a list of books, a wishlist of books, tagging them, and belonging to groups. Each user's activity, along with social network relationships, can thus be measured by several indicators. Figure 1 displays the probability distributions of these indicators. Some users do not use all the functionalities of the system. Not surprisingly, more than 90% of users share a list of books (of length $n_b$), as aNobii focuses on this aspect. Slightly more than 50% share a wishlist (length $n_w$), and also slightly more than 50% belong to at least one group ($n_g$ denotes the number of group memberships). Close to 3/4 have at least one friend or neighbor. In contrast, only about 30% of users use tags ($n_t$ distinct tags, and overall $a$ tagging events). Moreover, and as also observed in other systems [13], [14], all these distributions have broad tails, spanning several orders of magnitude, showing a strong heterogeneity between users' behaviors: for each activity type, no typical value of users' activity can be defined. Due to the low amount of tagging activity of aNobii users, we will in the remainder of this paper focus on the other activity patterns.

### B. Correlations and mixing patterns

As also investigated for Flickr and Last.fm in [14], [13], Figure 2 and Table II shows how the various activity measures are correlated, by displaying the average activity of users with $k_{out}$ out-neighbors in the social network, as measured by the various metrics defined above. For instance,

$$\langle n_b(k_{out}) \rangle = \frac{1}{|u : k_{u,out} = k_{out}|} \sum_{u : k_{u,out} = k_{out}} n_b(u)$$

gives the average library size of users with $k_{out}$ friends or neighbors. All types of activity show a clear increasing trend for increasing values of $k_{out}$, i.e. of the number of friends and neighbors that users have declared. The strong fluctuations visible for large $k_{out}$ values are due to the fewer highly-connected nodes over which the averages are performed. Interestingly, users with a large number of social contacts but with a low activity in terms of books or groups can be observed. Moreover (not shown), at any fixed value of $k_{out}$, strong fluctuations in $n_b$, $n_g$, $n_w$ are still present. Despite these important heterogeneities in the behavior of users with the same degree $k_{out}$, the data clearly indicate a strong correlation between the different types of activity metrics, as also shown by the Pearson correlation coefficients (table II).

We also note that the activity patterns are assortative in the social network: users tend to be linked with other users having similar activity patterns. This is measured for instance for the number of shared books, by measuring the average number of books of the friends and neighbors of users with a given $n_b$: this quantity displays an increasing trend as a function of $n_b$, showing that users who have many books on their library tend to be linked with other active users, while users with few books are linked with other less active users. Assortative mixing patterns have also been shown to exist in Last.fm and Flickr [14].

*C. Topical alignment*

Finally, we investigate the similarity of user profiles in relation to the social network structure. This can be measured for instance by the average number of common books in the libraries of pairs of users separated by a distance $d$ in the network, or by the corresponding average cosine similarity. The cosine similarity between libraries of users $u$ and $v$ is given by

$$\sigma_b(u,v) = \frac{\sum_b \delta_u(b)\delta_v(b)}{\sqrt{n_b(u)n_b(v)}}, \quad (1)$$

where the sum is over all possible books, $\delta_u(b)$ is 1 if user $u$ has book $b$ in the library, 0 otherwise, and $n_b(u)$ is the size of $u$'s library. As Fig. 3 shows, while the average number of shared books is quite large for neighbors, it drops rapidly as $d$ increases, and is close to 0 for $d \geq 5$ (similar results are obtained when measuring the similarity between users' groups memberships or tagging activities): the user profiles, in terms of shared exposed books, display a local alignment, which decays with the distance on the social network. Moreover, Fig. 3 also presents the results of the same measures of alignment for a null model in which books are randomly reshuffled between users, keeping for each user his number of books, and the statistics of the number of users having a given book, in the spirit of [14]. A comparison of the real data with the null model shows that part of the observed alignment is due to the assortative behavior of the users activities: active users tend to be connected to other active users, and it is statistically more probable to find common items between longer lists of books. It is nonetheless clear, both from the measure of the number of common books, and from the cosine similarity (which is normalized, hence insensitive to the activity of users), that the observed local alignment is a real effect. Connected users are more likely to have similar profiles: the presence of a social link is correlated with some degree of shared context between the connected users, which are likely to have some interests in common, or to share some experiences, or who are simply exposed to each other's activities.

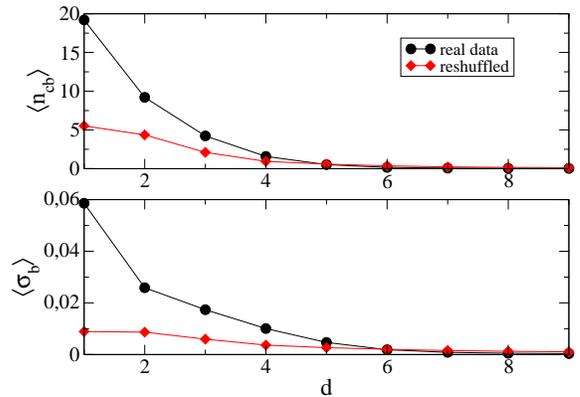

Fig. 3. Average similarity of the libraries of aNobii users as a function of their distance in the social network. The similarity is measured by the average number of common books (top, $\langle n_{cb} \rangle$), and by the average cosine similarity (bottom, $\langle \sigma_b \rangle$) between the book lists. In both cases data for the same social network with reshuffled booklists are shown.

The various properties of heterogeneous behaviors, correlations between activity measures, assortative mixing, and local alignment of profiles, have all been observed in other social networking systems [14]. Two important points have to be noted in this respect: as our database contains the complete largest connected component of aNobii, none of the sampling issues of other studies, which were dealing only with a small fraction of the social network, arise (for a thorough dissertation on sampling issues see [15]). Moreover, it is striking to observe that similar properties and regularities are observed across very diverse social networking systems, which have very distinct amounts of users, and address very different communities with very different interests.

V. GEOGRAPHIC ANALYSIS

Users can provide details about their geographical location. Location is composed by the country name plus an optional specification of the city. Since aNobii users are very prone to use this function (97% of accounts specify the country and 38% include also the city), an accurate investigation on geographic aspects and on their influence on social aggregation can be performed. The social network is split in two main geographic communities. The first is the Italian one, includes about 60% of the users; the latter, the Far East community, includes Taiwan (18%) and Hong Kong (10%). The proportions of other relevant countries are shown in Figure 4.

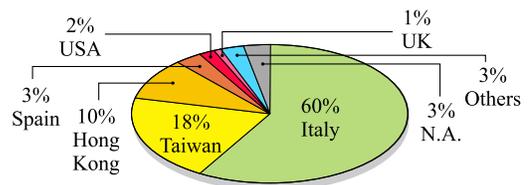

Fig. 4. Users nationalities in aNobii.

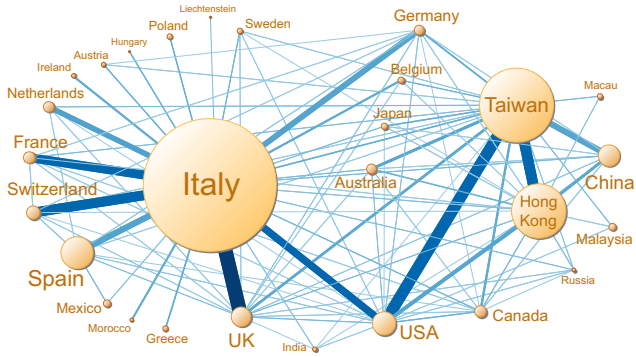

Fig. 5. Graph of aNobii countries. Nodes areas are scaled according to the size of the geographic communities and edges' width and colors are proportioned to the number of links that connects nodes between the countries. Small communities linked with the rest of the graph with less than 50 links are not represented.

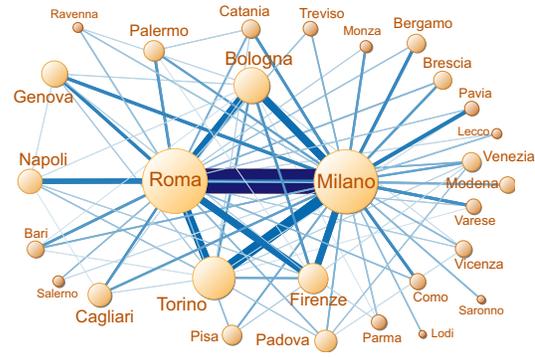

Fig. 6. Graph of aNobii Italian towns. Nodes areas and edges' width and colors are scaled like in Fig. 5. Small towns linked to others with less than 10 links are not represented.

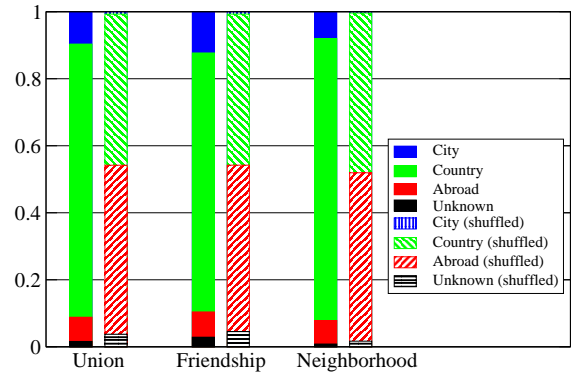

Fig. 7. Geographic homophily: fractions of possible geographical relationships between connected users (same city, same country, or users from different countries), for each network, compared with the same measure performed on shuffled networks.

Figure 5 depicts the aNobii social graph where nodes belonging to the same countries are clustered together. It clearly appears that the two main communities are loosely interconnected. As a result, the network is composed by two well-separated cores which are connected mostly by indirect bridges, for example through the mediation of the USA cluster. Conversely, intra-cluster connections are denser. For example, Figure 6 depicts a zoomed view on the Italy cluster, showing how clusters of users who live in the same city are interconnected. Of course, language is the first reason for the loose connection between national communities; since people tend to be interested in books written in their own mother tongue, social interactions with foreign users result to be infrequent. However, differences between languages alone do not entirely explain this phenomenon: for example, United Kingdom users are much more connected with Italians than with USA users.

In order to investigate more in detail this phenomenon, we checked how many links connect users that reside in the same country or city. Figure 7 shows that about 90% of the social ties link users in the same country, and in particular around 10% of links connect users who live in the same city. The remarkable fraction of connections between people from the same town is a signal of the presence of a geographic proximity homophily phenomenon in addition to language homophily. Users tend to connect with people who speak the same language, but the social aggregation process is also guided by a clear tendency to search for acquaintances that reside geographically close. To show that the observed homophily trends are not simply due to statistical properties, caused for example by the imbalance of the Italian cluster's size compared to others, we performed the same analysis on a reshuffled version of the social network, where each single user keeps its out-degree but rewires its links at random. The results, reported in Figure 7, show that ties connecting users from the same city almost disappear in the shuffled scenario and the inhomogeneity of countries of connected users rises significantly, slightly prevailing on homophily trend.

Finally, we extend the geographic homophily analysis to users at distance $d$ along the social network, in order to explore the relationship between geographic aggregation and distance in the social graph. Figure 8 shows the fraction of pairs of users with the same country or town at distance $d$ in the graph, compared with the null-model based on link reshuffling. In particular, in the real network we observe a high nationality homophily up to the fourth level of network proximity and a subsequent sudden fall, while the null-model curve starts from a value near $0.5$ and fades very slowly with the distance. A similar trend can be found for cities; here the real curve drops more rapidly, suggesting that people belonging to the same city tend to be connected directly. Such results mean that proximity in social network is closely related not only with users languages but also with geographic proximity.

The geographic homophily trend can be easily explained for the friendship network, because it is used mainly by people who already know each other in real life. However, the same tendency is found in the neighborhood network, which is basically a network of interests that is aimed to connect people with similar literary tastes. This is a new result compared to previous works that inspect spatiality in OSNs (e.g. [11]). We seized a hint of an inverse causal relationship: not only users search on OSNs people they know in real life yet, but among those who have a stronger affinity degree with them, they tend to choose people geographically close. A simple interpretation for this fact is that users want to know on the web people they would like to meet in real life too. Indeed, in the case of aNobii, groups of users frequently arrange literary-themed events.

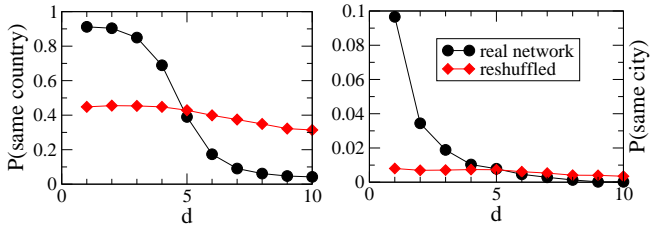

Fig. 8. Fraction of pairs of users at distance $d$ in the union network residing in the same country or town. In both cases data from the network with reshuffled links are shown.

| | $1 \to 2$ | $2 \to 3$ | $3 \to 4$ | $4 \to 5$ | $5 \to 6$ |
|---|---|---|---|---|---|
| New nodes | 2241 | 2121 | 1911 | 3214 | 3567 |
| Removed nodes | 239 | 222 | 230 | 220 | 684 |
| New edges | 19472 | 18324 | 17618 | 24805 | 26883 |
| Died edges | 642 | 763 | 713 | 782 | 700 |
| $u \to v$ | 5409 | 4942 | 5259 | 6546 | 6357 |
| Reciprocated | 1016 | 1155 | 1285 | 1526 | 1688 |
| $u \leftrightarrow v$ | 1809 | 1597 | 1604 | 1924 | 2235 |
| Simple closure | 2070 | 1976 | 2143 | 2497 | 2382 |
| Double closure | 955 | 904 | 877 | 1027 | 1141 |

TABLE III
EVOLUTION OF SOME QUANTITIES FROM ONE SNAPSHOT TO THE NEXT.

## VI. DYNAMICAL ANALYSIS AND NETWORK GROWTH

In this section, we analyze the temporal evolution of the social networking system under consideration. Our dataset contains indeed the time evolution for the whole largest connected component of the aNobii social network during 2 and a half months, providing the opportunity to investigate how new users behave, and how new links between users are created: while the study of a given snapshot allows for instance to grasp correlations between the distance on the network and the users' profiles similarity, investigating the time evolution can give hints about causality.

### A. Triangle closure and preferential attachment

Table III gives the evolution of some basic network quantities when comparing one snapshot to the next. New users are constantly joining the social network's largest component, and create edges towards already present users. Very few users leave the network, and very few edges disappear between nodes which remain in the network ("Died edges"). Edges on the other hand are created between already present users. We classify these nodes in various categories. $u \to v$ denotes the number of new non-reciprocal links, and $u \leftrightarrow v$ the number of new reciprocal links, between users already present in the first snapshot and which were not yet connected. "Reciprocated" gives instead the number of new links from a user $u$ to a user $v$, such that a link from $v$ to $u$ already existed. "Simple closure" and "Double closure" refer respectively to links of the type $u \to v$ and $u \leftrightarrow v$ which close at least a directed triangle (for instance, $u \to w \to v$ existed in the first snapshot): in a social network, one can indeed expect to see the creation of links towards a "friend of a friend". Table III makes clear that the aNobii social network's largest connected component is in constant growth, as very few nodes and edges disappear. Users moreover update their activities. Despite this evolution, the statistical features shown in Sections IV and V are stable over time.

We now focus more specifically on the way new users join the network's largest component, and new links are created. We first study the properties of the users *towards which* new users create links. In particular, we test in Fig. 9 (top) the hypothesis that a preferential attachment mechanism is at work, such that users with already large numbers of links are preferentially chosen [16]. The method to quantify preferential attachment is as follows [17]: let us denote by $T_k$ the a priori probability for a newcomer to create a link towards a node of degree $k$, between time $t-1$ and $t$. Given that at time $t-1$ the degree distribution of the $N(t-1)$ nodes is $P(k, t-1)$ (i.e., there are $N(t-1)P(k,t-1)$ nodes of degree $k$), the effective probability to observe a new link from a new node to a node of degree $k$ between $t-1$ and $t$ is $T_k P(k, t-1)$. We can therefore measure $T_k$ by counting for each $k$ the fraction of links created by new nodes that reach nodes of degree $k$, and dividing by the measured $P(k, t-1)$. As shown in Figure 9, we obtain a linear behavior, both when considering for $k$ the in and the out-degree (which are strongly correlated). This is a clear signal of a linear preferential attachment. As new users clearly do not have a knowledge of the network structure at $t - 1$, and do not have a priori a particular motivation to create links towards highly connected other users, it seems plausible to infer that this effective preferential attachment arises from the fact that a new user will create a link not only towards another user (who he may know for other reasons) but also towards some of this user's neighbors. Such locally driven mechanism is indeed known to result in effective preferential attachment [18], [19], and we have checked that it is indeed present in our case: many users join the network's largest component by creating links to pairs of already connected users.

### B. Causal connection between similarity and link creation

Let us now consider the links created between two snapshots $t_0$ and $t_0+1$ by users who are already present in the snapshot $t_0$. As Table III shows, a large part of these new links close triangles, i.e., connect neighbors of neighbors: these new links connect users who were at distance $d = 2$ on the social network at $t_0$. Moreover, Fig. 9 (bottom) shows that the distribution of distances at $t_0$ for the pairs of nodes which become linked between $t_0$ and $t_0 + 1$ is strikingly different from the global distribution of distances along the social graph, and biased towards small distances. This is also linked with the fact that most new links connect users in the same country (in proportions similar to the ones of Fig. 7). The fact that most new links connect nodes which were already close on the social network, together with the decrease of the similarity of users' profiles as a function of the distance on the social network, has as obvious consequence that the average similarity between nodes that become linked between $t_0$ and $t_0 + 1$ will be larger than the average similarity of random pairs of nodes. We compare therefore the average similarity between pairs of users which are at distance $d = 2$ at $t_0$ with the average similarity, also at $t_0$, between pairs of users who become linked between $t_0$ and $t_0 + 1$. Strikingly, the latter is larger than the former for all measures of similarities, as shown in Table IV, especially for bidirectionally created links, and links which close triangles. The probability that two users at distance 2 have 0 similarity is also much smaller for users who become linked between $t_0$ and $t_0 + 1$.

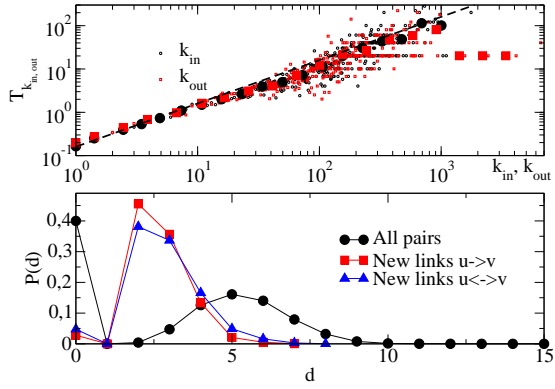

Fig. 9. Top: Measure of the preferential attachment. The dashed line gives a linear relationship. Bottom: distribution for the snapshot number $t_0 = 4$ of the distances of nodes which become linked between snapshots $t_0$ and $t_0 + 1$, compared with the distribution (for snapshot $t_0$) of distances between all pairs of users. $d = 0$ corresponds in fact to $d = \infty$, i.e. the points in $d = 0$ give the percentage of ordered pairs of nodes between which no (directed) path exists.

|  | $\langle n_{cb} \rangle$ | $\sigma_b$ | $\langle n_{cg} \rangle$ | $\sigma_g$ |
|---|---|---|---|---|
| $d_{uv} = 2$ | 9.5 (0.2) | 0.02 | 1.12 (0.61) | 0.05 |
| $u \to v$ | 12.9 (0.16) | 0.04 | 1.1 (0.6) | 0.08 |
| $u \leftrightarrow v$ | 18.5 (0.06) | 0.04 | 1.67 (0.44) | 0.11 |
| Simple closure | 18.2 (0.09) | 0.04 | 1.81 (0.45) | 0.1 |
| Double closure | 23.4 (0.03) | 0.05 | 2.2 (0.36) | 0.12 |

TABLE IV
AVERAGE SIMILARITY FOR SNAPSHOT $t_0 = 4$ OF PAIRS FORMING NEW LINKS BETWEEN $t_0$ AND $t_0 + 1$, COMPARED WITH THE AVERAGE SIMILARITY OF ALL PAIRS AT DISTANCE $2$ AT $t_0$. THE SIMILARITY IS MEASURED BY THE NUMBER OF COMMON BOOKS $n_{cb}$ OR GROUPS $n_{cg}$, AND BY THE CORRESPONDING COSINE SIMILARITIES $\sigma_b$ AND $\sigma_g$. THE NUMBERS IN PARENTHESIS GIVE THE PROBABILITY TO HAVE SIMILARITY EQUAL TO 0.

The picture emerging is thus the following: new links connect users who were already close, very often neighbors of neighbors; moreover, these users had more similar profiles than the average pairs of users at distance 2. In this respect, one can infer a first causal effect, namely that similarity of users partly drives the creation of new links.

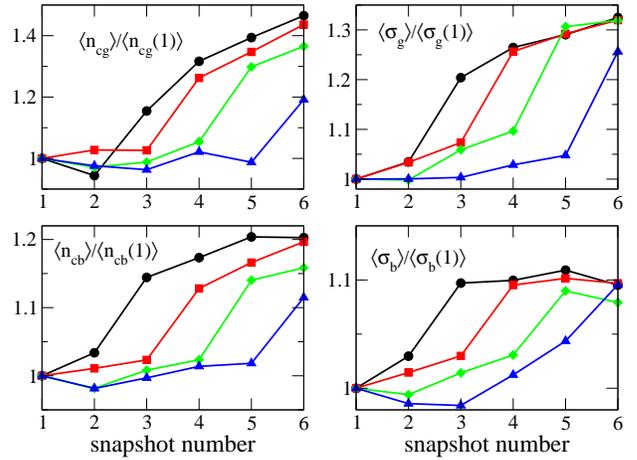

Fig. 10. Evolution of the average similarity of users' profiles, as measured by the numbers of common books or groups, and by the cosine similarities, from the first to the last snapshot, for links created between $t_0$ and $t_0 + 1$, for $t_0 = 2$ (black circles), 3 (red squares), 4 (green diamonds), 5 (blue triangles), normalized by the average similarity in the first snapshot. Similarities are rather stationary before $t_0$, and clear jumps are observed between $t_0$ and $t_0 + 1$.

We also investigate in Fig. 10 the time evolution of the average similarity of users' profiles, for libraries and group memberships, for pairs of users forming new bidirectional links between $t_0$ and $t_0 + 1$, for various $t_0$. Before the creation of the links, the similarity is stationary; a large jump is observed when the links are created, and the similarity continues to grow, albeit at a slower pace, after the link formation. This result strongly hints at the following scenario: after users create links, they take inspiration from their new neighbors for new books to read and new groups to join, and as a consequence align their profiles.

In a nutshell, our analysis on the dynamics of social aggregations reveals the presence of a *bidirectional* causal relationship between user similarity and social connections. Indeed, higher similarity determines a higher connection probability and, on the other side, users who get linked get more similar due to the significant influence that new acquaintances exert on one another. These results show that the mutual influence between social aggregation and user similarity holds not only for collaboration networks [9], but also for the general case of interest-based networks such as aNobii, where the similarity between users is evaluated on the basis of chosen and shared items, shared metadata, and topics of interest.

## VII. CONCLUSIONS

In this paper we studied structural and evolutionary features which are crucial to clear up the link formation process in Online Social Networks, using an empirical set of temporal snapshots of a fairly complete portion of the aNobii social network. We inspected topological aspects of the network by relating them with features that describe users, like books they have read or thematic groups they belong to. We observed that the creation of a social tie is strongly driven by *homophily* and *proximity*. In addition to being strongly influenced by language barriers, users tend to establish social ties with people with similar interest and which are near both in terms of social network hops and of geographic distance. We verified that this trend applies in a geographically driven network, where ties are established between users who know each other in real life, as well as in a pure network of interest.

The geometry of link creation reveals that *reciprocation* and *triadic closure* are very common patterns in the social graph evolution. A clear tendency to preferential attachment is observed in the process of addition of new users to the largest network component; this arises because of a process of *imitation* that leads users to connect to pairs or groups of connected users.

Finally, we investigated the causal relationship between the high similarity of users' profiles, in terms of books read and group membership, and the creation of a link between them. We found a robust empirical evidence of a *reciprocal* causal relationship: users search for friends with similar tastes and, once they are linked, a clear reciprocal influence leading to a stronger *profile alignment* occurs.

Since tags are not widely used in aNobii, we could not verify if the same phenomenon results also in vocabulary alignment. Reproducing the same analysis on OSNs with a higher fraction of taggers could be interesting to answer this question. Besides, a longer network monitoring could reveal if the profile alignment phenomenon remains stable over time. Furthermore, our findings could be profitably applied to link prediction tasks based on users features.